\documentclass{research4cacm}

\usepackage{cite,hyperref}

\usepackage{color}
\usepackage{verbatim}
\graphicspath{{Figs/}}
\hypersetup{hidelinks}
\usepackage{epstopdf}

\begin{document}

\title{Belt and Braces: When Federated Learning Meets Differential Privacy
\thanks{This article has been accepted by and is to appear in Communications of the ACM (CACM).}
}

\numberofauthors{5}
\author{
\alignauthor
Xuebin Ren %\titlenote{X. Ren is an associate professor in National Engineering Laboratory for Big Data Analytics, at Xi'an Jiaotong University, P.R. China.}
\\
%\affaddr{National Engineering Laboratory for Big Data Analytics}\\
\affaddr{Xi'an Jiaotong University}\\
\email{\normalsize xuebinren@mail.xjtu.edu.cn}
\alignauthor
Shusen Yang %\titlenote{S. Yang (corresponding author) is a professor in National Engineering Laboratory for Big Data Analytics, at Xi'an Jiaotong University, P.R. China.}
\\
%\affaddr{National Engineering Laboratory for Big Data Analytics}\\
\affaddr{Xi'an Jiaotong University}\\
\email{\normalsize shusenyang@mail.xjtu.edu.cn}
\alignauthor
Cong Zhao %\titlenote{C. Zhao is an associate professor in National Engineering Laboratory for Big Data Analytics, at Xi'an Jiaotong University, P.R. China.}
\\
%\affaddr{~~~}\\
%\affaddr{Department of Computing}\\
%\affaddr{~~~}\\
\affaddr{Xi'an Jiaotong University}\\
\email{\normalsize congzhao@xjtu.edu.cn}
\and
\alignauthor
Julie McCann %\titlenote{J. McCann is a professor in Department of Computing at Imperial College London, London, U.K.}
\\
%\affaddr{~~~}\\
%\affaddr{Department of Computing}\\
%\affaddr{~~~}\\
\affaddr{Imperial College London}\\
\email{\normalsize j.mccann@imperial.ac.uk}
\alignauthor
Zongben Xu %\titlenote{Z. Xu is a professor in National Engineering Laboratory for Big Data Analytics, at Xi'an Jiaotong University, P.R. China.}
\\
%\affaddr{National Engineering Laboratory for Big Data Analytics}\\
\affaddr{Xi'an Jiaotong University}\\
\email{\normalsize zbxu@mail.xjtu.edu.cn}
}

\maketitle

\begin{abstract}
\textcolor{black}{Federated learning (FL) has great potential for large-scale machine learning (ML) without exposing raw data.
Differential privacy (DP) is the \textit{de facto} standard of privacy protection with provable guarantees.
Advances in ML suggest that DP would be a perfect fit for FL with comprehensive privacy preservation. Hence, extensive efforts have been devoted to achieving practically usable FL with DP, which however is still challenging.
Practitioners often not only are not fully aware of its development and categorization, but also face a hard choice between privacy and utility. 
Therefore, it calls for a holistic review of current advances and an investigation on the challenges and opportunities for highly usable FL systems with a DP guarantee.}
\textcolor{black}{In this article, we first introduce the primary concepts of FL and DP, and highlight the benefits of integration. We then review the current developments by categorizing different paradigms and notions. Aiming at usable FL with DP, we present the optimization principles to seek a better tradeoff between model utility and privacy loss. Finally, we discuss future challenges in the emergent areas and relevant research topics.}
\end{abstract}

\section{Introduction}\label{sec:introduction}\vspace{3mm}

With the development of advanced algorithms, computing capabilities, and available datasets, machine learning (ML) have been widely adopted to solve real-world problems in various application domains. 
The success of ML often relies on large amounts of application-specified training data\textcolor{black}{, especially for large models like ChatGPT.} However, these data are often generated and scattered among enormous network edges or users' end devices, and can be quite sensitive and impractical to be moved to a central location as the result of regulatory laws (e.g., GDPR) or privacy concerns~\cite{cheng2020federated}. This fact has brought an inconvenient dilemma between large-scale ML and increasingly severe data isolation. The conflict between data hungriness and privacy awareness is becoming increasingly prominent \textcolor{black}{in the artificial intelligence (AI) era}. 

Google proposed FL as a potential solution to the above issue~\cite{mcmahan2017communication}. Through coordination between the central server and clients (devices participated in FL), FL collaboratively trains ML models over extensive data across geographies, which bridges up the gap between an ideal of big data utilization and the reality of data fragmentation everywhere.
\textcolor{black}{By sharing locally trained models, FL not only minimizes the risks of raw data exposure but also eliminates the client-server communications.} 
\textcolor{black}{Once proposed, it has been seen as a rising star in AI technology. Its recent usage in fine-tuning of large language models (LLMs) confirmed that again.}

The advancement of FL in privacy protection stems from the delicacy in restricting raw data sharing.
\textcolor{black}{This is however far from sufficient, as gradients of deep models can even expose the privacy~\cite{zhu2019deep} but FL gives no formal privacy guarantees.} 
Fortunately, \textcolor{black}{differential privacy (DP), proposed by Dwork, allows controllable privacy guarantee, via formalizing the information derived from private data~\cite{Dwork2011Firm}. 
By adding proper noise, DP guarantees a query result does not disclose much information about the data. 
Because of its rigorous formulation, DP has been the \textit{de facto} standard of privacy and applied in both ML and FL.}

As privacy in design, \textcolor{black}{the emergence of DP and FL greatly encourages data sharing and utilization in reality.} 
On one hand, \textcolor{black}{by restricting raw data exposure}, FL enables ML model training over massively fragmented data. It also significantly enriches ML applications for extensive distributed scenarios. 
On the other hand, by rigorously limiting the indirect information leakage, DP can strengthen the privacy in trained models with provable guarantees.
The complementarity of FL and DP in privacy suggests a promising future of their combination, which can significantly extend the applicable areas for both techniques and bring privacy-preserving large-scale ML to reality.
Specifically, FL has advantages in fusing geographically isolated datasets, while DP can offer provable guarantees and thus encourage sensitive data sharing. 
Aimed at exploiting the potential of ML to its fullest, it is highly desirable and essential to build FL with DP to train and refine ML models with more comprehensive datasets.

\textcolor{black}{The benefit of privacy protection in both FL and DP comes at a cost in terms of data utility, albeit other issues. 
FL clients often have limited capabilities and distribution-skewed datasets, causing insufficient and/or unbalanced training of global models with low utility. 
DP algorithms hide the presence of any individual sample or client by adding noise to model parameters, also leading to possible utility loss. 
Therefore, utility optimization, i.e., improving the model utility as much utility as possible for a given privacy guarantee is an essential problem in the combining use of FL and DP.  
Given the great potential, studies on this problem have rapidly expanded in recent years.
However, they are often conducted based on various FL and DP paradigms concerning different security assumptions (e.g., whether the server is trustworthy) and levels of privacy granularity (e.g., sample or client). 
Without a systematic review and clear categorization of existing paradigms, it is hard to precisely evaluate and compare their utility performance. 
On the other hand, despite the paradigm differences, the utility optimization principles are quite similar. However, current studies often focus on specific algorithm design for different paradigms of FL with DP and there lacks some common pathways to follow. Meanwhile, the only few surveys on the intersection of DP and FL either have different focus other than the utility issue or lack  high-level insights into the future challenges.
}

\textcolor{black}{Here, this article aims to provide a systematic overview of DP-enabled FL while focusing on high-level perspectives on its utility optimization techniques.
We begin by presenting an introduction to FL and DP respectively, highlighting the benefits of their combination. We then summarize research advances by categorizing the paradigms and software frameworks of FL with DP. Aiming at usable analytic results, we present the high-level principles and primary technical challenges in their utility optimization in several emerging scenarios. Finally, we discuss some related topics to FL with DP, which would also impact the achieved data utility. 
Our review can benefit the general audience with a systematic understanding of the development and achievements on this topic. The perspectives on utility optimization for DP-enabled FL can offer some insights into research opportunities and challenges for usable AI services with privacy protection in both academia and industry.}

\section{Federated Learning}\vspace{3mm}

\subsection{Overview of Federated Learning}\vspace{3mm}

An FL system is essentially a distributed ML (or DML) system coordinated by a central server~\footnote{\small Decentralized FL is a special form where clients collaborate via peer-to-peer communication without a server.}, which helps multiple remote clients with separate datasets to collaboratively train an ML model, under a privacy constraint that any client does not expose its raw data. 
There are two popular FL frameworks~\cite{mcmahan2017communication}. Federated stochastic gradient descent (FedSGD) is the federated version of the stochastic gradient descent (SGD) algorithm. In SGD for centralized ML, gradients are computed on a random subset of the total dataset and then used to make one step of the gradient descent. FedSGD uses a random fraction of clients and all their local data. The gradients are averaged by the server proportionally to the number of training samples on each client and used to make a gradient descent step. To overcome the communication bottleneck, federated averaging (FedAvg) allows clients to perform more than one batch update on the local dataset and exchange the updated parameters rather than the gradients~\cite{konevcny2016federated}. FedAvg is a generalization of FedSGD since averaging the gradients would be equivalent to averaging the parameters themselves if all the clients begin with the same initialization.
So, generally FL works as follows:
1) Each participating client performs a local training procedure on its own dataset and sends the gradients or model updates to the server.
2) The server securely aggregates the received gradients or model updates, and updates the global model accordingly.
3) The server sends back the new global model to the corresponding clients.
4) The clients update their local models and prepare for the next iteration.
The above procedures are repeated until the global model converges or a sufficient number of iterations are applied.
FL is classified into cross-device FL that leverages up to millions of devices in the wide-area network, and cross-silo FL that ties up a handful of edge nodes with reliable backbones.

\subsection{Comparison with Traditional DML}\vspace{3mm}

\begin{figure}[tbp]
	\centering
	{
		\includegraphics[width=0.45\textwidth, height=135pt]{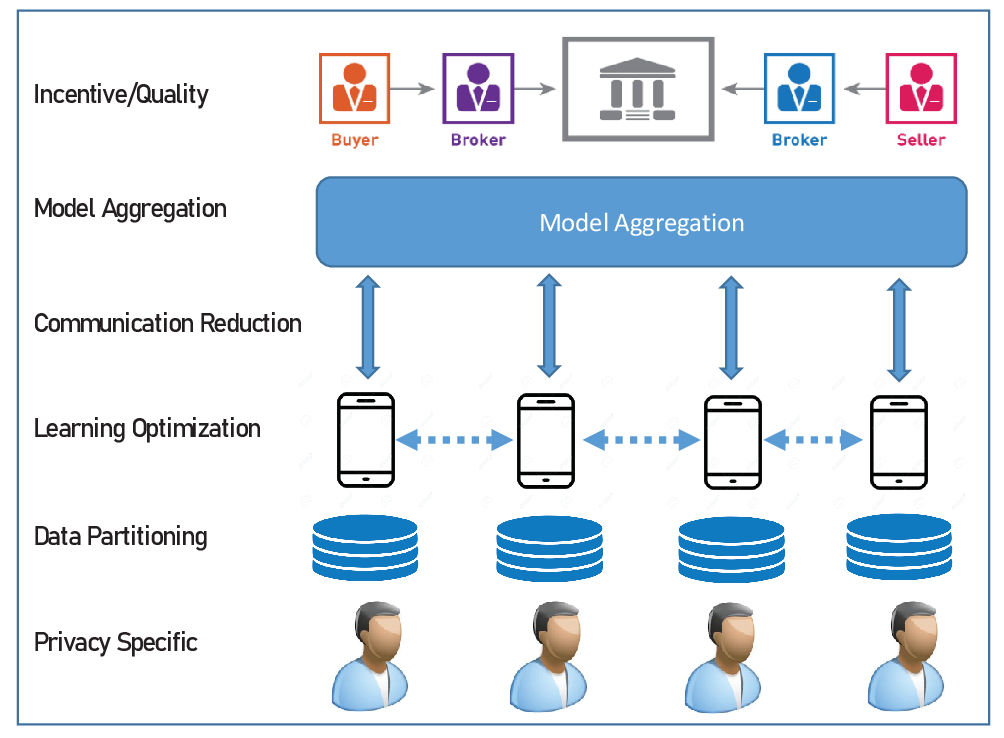}
	}
	\center\caption{Building Blocks of FL Systems \label{fig: buildingblocks}}
\end{figure}

Despite being a typical DML paradigm, when compared with \textit{traditional DML in data centers} for ML speedup, FL has many distinct characteristics (as shown in Fig.~\ref{fig: buildingblocks}):

\textbf{Privacy requirement}:
Unlike traditional DML in the data centers (where data can be arbitrarily scheduled among computing nodes), ensuring privacy protection lies at the center of FL, which strictly prohibits raw data sharing.
	
\textbf{Data partitioning}: Data in FL are generated naturally or obtained from individual users, thus often being non-IID and imbalanced. Instead, data in traditional DML are usually manually scheduled to be almost shuffled or balanced.

\textbf{On-device learning:} In data centers, DML computing nodes are homogeneous, deployed centrally, and powerful. In contrast, FL is implemented with tens to millions of distributed clients with heterogeneous and limited computing capacities.

\textbf{Communication:}
Traditional DML in data centers can enjoy Gigabytes bandwidth and communicate in a peer-to-peer manner. However, FL clients are usually connected to the server by the wide-area network and bandwidth constrained.

\textbf{Model aggregation:} 
Model aggregation fuses training results (e.g., local models) from distributed nodes.
Compared to homogeneous sub-models in traditional DML, one challenge in FL is the prominent heterogeneity among local models due to either non-IIDness or varied training progresses.

\textbf{System actors:} 
Unlike the closed and fixed system of traditional DML, FL is often conceived as an open and scalable system consisting of massive clients owned by different individuals/organizations seeking for different benefits. 

\subsection{Privacy Threats in Federated Learning}\vspace{3mm}
Due to above characteristics, e.g., geographically distribut-ed nature, open architecture, and complicated interactions \cite{RigakiGarcia-45,wang2019beyond,nasr2019comprehensive}, various attack can be mounted against FL in both model training and serving (i.e., inference). 
Instead of those for degrading system availability or compromising data integrity (e.g., poisoning attacks), we focus on privacy threats for snooping private information in FL.

\textbf{Privacy Adversaries}.
Privacy may be disclosed to or inferred by anyone that has access to the information flow in FL.  
Compared with ML over centralized data or traditional DML centrally deployed in datacenters, mutually distrusted entities in FL may all be viewed as privacy adversaries inferring others private information
The possible adversaries can be classified as insiders and outsiders. The former includes the server and participating clients, and the latter contains eavesdroppers over communication channels and third-party analysts (users) consume the final model.
Compared with the outsiders that are more likely to have black-box access (i.e., can only query via APIs) to the final model, insiders are generally more capable as they can often have white-box access (i.e., full access with prior knowledge) and substantially impact FL model training. The insiders can be further considered to be semi-honest and malicious. The former is also known as honest-but-curious, i.e., following the protocol correctly but tries to learn other entities' private state. The latter may actively deviate from the protocol (e.g., modifying data or colluding with others) to achieve the goal.

\begin{figure}[tbp]
	\centering
	{
		\includegraphics[width=0.5\textwidth]{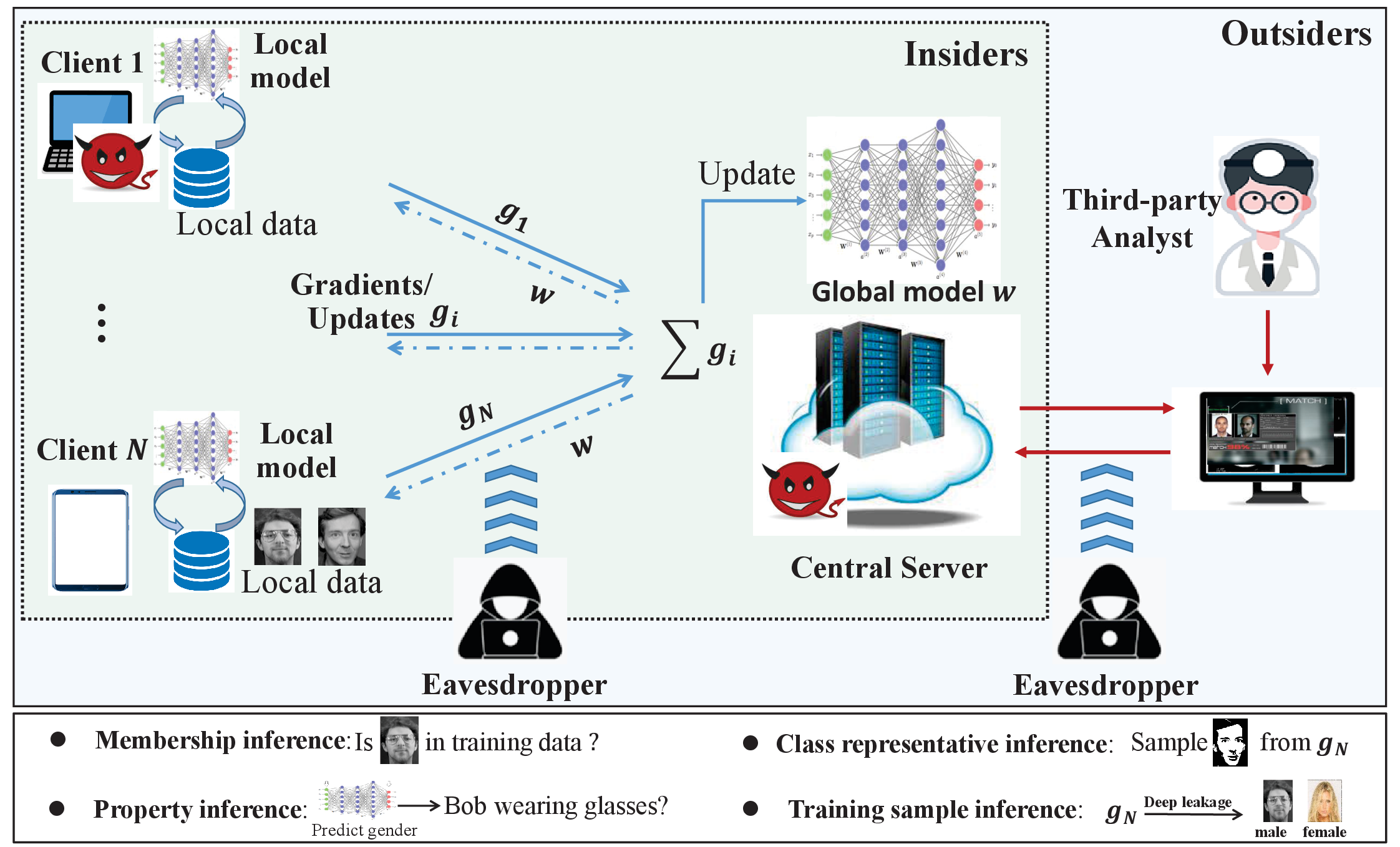}
	}
	\center\caption{Privacy threats in FL training \label{fig: architecutere}}
\end{figure}

\textbf{Privacy Attacks}.
Considering above adversaries, the following privacy attacks may exist in FL (shown in Fig.~\ref{fig: architecutere}):

\textit{Membership inference} targeting a model aims to predict whether a given data sample was in its training set~\cite{shokri2017mia}. It works by training multiple customized inference models to recognize noticeable patterns in the models' outputs for the given sample.
In traditional ML centrally deployed, membership inference is normally mounted by third-party users. 
In FL, it can be carried out by not only third-party users, but also communication eavesdroppers, and even participating clients and the server~\cite{nasr2019comprehensive}. This is because, the local, aggregated, accumulated and final forms of gradients or model parameters, all may expose private information about training data~\cite{melis2019exploiting}.
Moreover, active attackers disguised as clients can selectively alter their gradient updates to significantly enhance the attack accuracy over the victim clients~\cite{RigakiGarcia-45}.

\textit{Class representative inference} tries to generate class representatives from the underlying distribution of the training data that the targeted model could have been trained on. In traditional ML, third-party users can achieve this goal by iteratively modifying the features of a random sample until a maximal confidence reaches~\cite{fredrikson2015inversion}, or training an inverse model, with black-box access to the targeted model. In FL, while a honest-but-curious server may partially recover some samples of honest clients by simply observing their uploaded gradients,
active malicious clients or a passive malicious server can exploit generative adversarial networks (GANs) to construct class representatives from not only the global data distribution but also specific clients~\cite{hitaj2017deep}. 

\textit{Other privacy attacks} include inferences for properties, and even the accurate training data (both inputs and labels). Different from above inferences in terms of properties characterizing an entire class, property inferences aim to infer those properties independent of the characteristic features. With some auxiliary data, a passive adversary trains a binary property classifier to predict whether the observed updates were based on the data with the property, while an active adversary can exploit multi-task learning to simultaneously conduct main FL training and infer the targeted property state with enhanced capability. Inferring accurate training data is also demonstrated possible under the \textit{deep leakage from gradient}, which optimizes the dummy inputs and labels via minimizing the difference between the dummy and targeted gradients for differentiable models~\cite{zhu2019deep}.

\subsection{Related Privacy-preserving Techniques}\vspace{3mm}
\label{sec: secure technique}

Cryptographic primitives and protocols, can restrict unauthorized access to confidential information, thus reducing the chances of privacy leakage~\cite{kairouz2019advances}. For instance, homomorphic encryption (HE) supports dedicated operations on multiple encrypted data to produce ciphertexts that can be decrypted to generate desirable functional outcomes of original plaintexts. 
\textcolor{black}{Functional encryption (FE) authorizes the holder of a key associated with a specified function to directly learn the function output over encrypted data and nothing else.}
Using secure multi-party computation (SMC), a set of parties jointly compute from their inputs without relying on a trusted third party or learning each other's input. 
\textcolor{black}{
Cryptography implemented in software still requires error-free environment for execution and uncompromising storage of secret key. This naturally calls for hardware-assisted security. Trusted execution environments (TEEs) can create an isolated operating environment that ensures the confidentiality of the data and codes within, while enabling remote authentication and attestation.} 
\textcolor{black}{In FL training, above technologies can be adopted either alone or in combination to guarantee desired confidentiality of the processed models.}

\textcolor{black}{However, note that privacy is essentially orthogonal to confidentiality. Whatever secure protocols and trusted systems are used, a final model will eventually be trained for consumption. Even if providing inference APIs only, model predictions may still reveal sensitive information as ML models inevitably carry some knowledge of training samples~\cite{dwork2014algorithmic}.}
In general, models with poor generalization tend to leak more. Overfitting is one of the sufficient conditions of performing membership inference attacks~\cite{RigakiGarcia-45}.
Therefore, another line of defensive approaches is properly suppressing fine-grained model utility. For instance, regularization can undermine inference attacks by reducing overfitting. For deep learning, two useful strategies are model compression (or sparsification) that sets gradients below a threshold to zero and weight quantization that limits the parameter precision. However, these approaches provide intuitive protection only without rigorous guarantee~\cite{shokri2015privacy}.

\section{Differential Privacy}\label{sec: DP}\vspace{3mm}

With provable guarantee of limiting privacy leakage even in securely aggregated results, differential privacy is promising to complement above technologies and strengthen FL.

\subsection{Overview of Differential Privacy}\label{sec: DP overview}\vspace{3mm}
Through establishing a formal measure of privacy loss, DP allows rigorously controlling the (worst-case) information leakage. 
Informally, it guarantees an algorithm's output does not change much for two datasets differing by a single entry~\cite{Dwork2011Firm}. To achieve DP, the basic idea is to properly randomize the relationship between data input and algorithmic output, e.g., by adding noise. 

DP has various models, as noise can be added to the different components or phases of algorithms~\cite{dwork2014algorithmic}.
Conventional DP assumes a trustworthy aggregator and adds minor noise to algorithm output, which is known as centralized DP (CDP). Assuming an honest-but-curious aggregator, local DP (LDP) randomizes data at users' end before collection, and reconstructs utility from perturbed data of multiple uses. 
From CDP to LDP, the trust model is weakened under the same DP parameter, while data uncertainty and accuracy loss becomes larger. To bridge the trust-accuracy gap, distributed DP (DDP) exploits cryptography to obtain high accuracy without a trusted aggregator~\cite{wagh2021dp}. There are currently two DDP paradigms, based on secure shuffling and secure aggregation respectively. Secure shuffling uses an anonymous communication channel to alleviate identification risks of messages and thereby relaxing the trust model~\cite{cheu2019distributed}. Secure aggregation replaces the trusted aggregator by secure computation protocols and thus can reduce noise and gain the same utility as in centralized model.

The prevalence of DP also comes from many delicate characteristics~\cite{dwork2014algorithmic}. The post-processing property keeps the privacy guarantee of algorithms after arbitrary workflows. Composition theorems help to understand the composed privacy guarantee of a series of sub-algorithms and enables building complicated algorithms from simple operations. 

\begin{figure}[tbp]
	\centering
	{
		\includegraphics[width=0.4\textwidth, height=140pt]{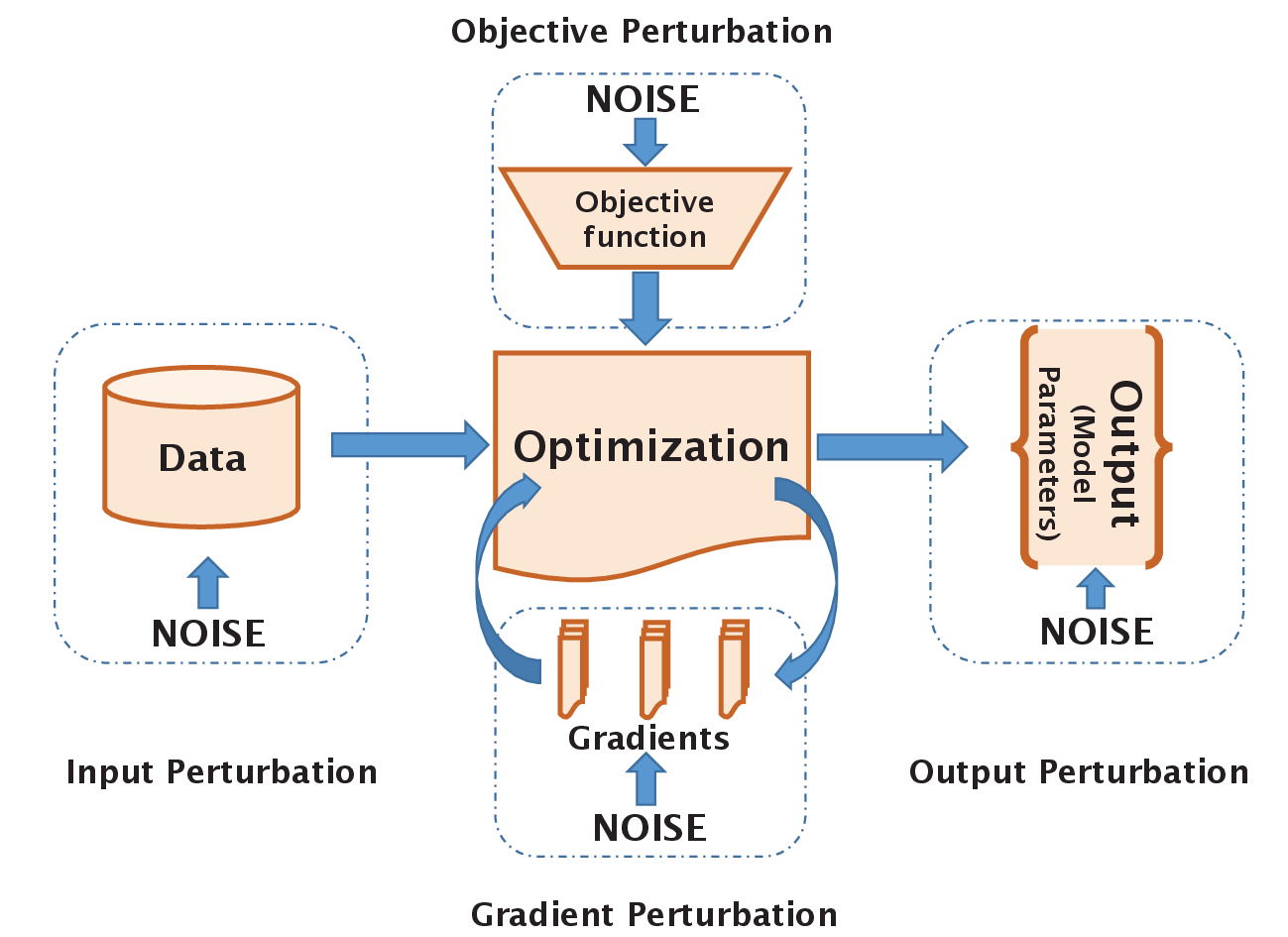}
	}
	\center\caption{Approaches to achieve DP for ML \label{fig: dpml}}

\end{figure}

\subsection{Differential Privacy for ML}\label{sec: DP for ML}\vspace{3mm}
DP has been applied in ML to prevent adversaries with access to the model from inferring the training data~\cite{ji2014differential}. Except that \textit{intrinsic privacy} can be achieved freely for some ML models with inner randomness~\cite{hyland2019intrinsic}, noise addition to different components of ML algorithms provides viable pathways for privacy-preserving ML with DP, as shown in Fig.~\ref{fig: dpml}.

\textit{Output perturbation} adds calibrated noise to the parameters of final models~\cite{chaudhuri2009privacy},
which, however, may have large (even unbounded) sensitivities and lead to severe model utility loss. 
\textit{Input perturbation} randomizes training data and then constructs an approximate learning model on it~\cite{farokhi2020distributionally}. 
Similar to LDP, it has learning limits and low model utility~\cite{duchi2014privacy}.
\textit{Objective perturbation} perturbs the objective functions of the optimization problem in ML. Although functional mechanism~\cite{zhang2012functional} allows its usage for complicated model functions, it is often infeasible to explicitly express the loss functions for most ML models, especially deep learning. 
\textit{Gradient perturbation} that sanitizes parameter gradients during training~\cite{shokri2015privacy} can ensure DP even for nonconvex objectives, making it much useful for deep models. 
Differentially private SGD (DP-SGD) has now been the common practice for privacy-preserving ML~\cite{abadi2016deep}. It works by sampling a min-batch of samples, clipping the $l_2$ norm of the gradients computed on each sample, aggregating the clipped gradients, and adding Gaussian noise in each iteration. By incorporating gradient clipping, it can avoid the issue of unknown gradient sensitivity. Besides, it is often used with moments accountant for tracking a tighter privacy loss bound.

\section{Federated Learning with Differential Privacy}\vspace{3mm}
The wide application of DP in privacy-preserving ML shows the great potential of privacy-preserving FL with DP.

\subsection{Benefits of FL with DP}\label{sec: DP Benefits}\vspace{3mm}

DP with rigorous guarantee has been an essential technology for privacy-preserving data analysis and ML. Although it has been successfully integrated into distributed systems for data querying and analyses~\cite{roy2020crypt,bater2018shrinkwrap}, there is still a lack of DP-enhanced framework for large-scale distributed ML over massively scattered datasets. 
FL supports flexible ML tasks with extensive models and scalable ML training for massively scattered datasets. Despite ensuring no direct data exposure by solely sharing intermediate parameters, it still lacks a formal privacy guarantee and may expose indirect privacy. 
Therefore, when combining them together, FL with DP can realize large-scale and flexible distributed learning while preventing both direct and indirect privacy leakage.

As complements of each other towards the same goal of encouraging massively confidential and sensitive data utilization, the combination of FL and DP can achieve paramount benefits for privacy protection in reliability.
\textbf{FL empowers and prospers DP-based ML over large-scale siloed datasets.} 
DP-based ML (especially deep learning) in the centralized setting, has made a rapid progress. However, data centralization and privacy regulations strongly hinders its further development.
As a result, DP-based ML wishes to meet large-scale data or data-extensive applications. 
Fortunately, FL naturally enables DP-based ML over massively scattered data, thus greatly prospering its success. 
\textbf{DP completes and strengthens the reliability of FL via offering rigorous guarantee.} 
The mission of FL is to train and refine ML models with more comprehensive end-user data, which is subject to the willingness of data owners. 
Hence, provable privacy guarantee is key to the popularization of FL systems. 
Beyond isolated datasets, privacy-preserved FL systems may encourage users to contribute more sensitive datasets.

\subsection{Research Advances on FL with DP}\label{sec: related of FL DP}\vspace{3mm}
Due to above benefits, marrying FL with DP has attracted extensive interests from both the academia and industry.
We systematically review the advances according to different paradigms and privacy notions.

\subsubsection{FL with Centralized DP} \vspace{3mm}
It is natural to extend differentially private ML algorithms (e.g., DP-SGD) in centralized setting, to the context of FL to prevent information leakage from the training iterations and final model, against malicious clients or third-party users.
	
DP has different granularity, relying on the precise definition of neighboring datasets. 
Different from DP-SGD that provides \textit{sample-level DP} for hiding the existence of any single sample, it is more meaningful to provide \textit{client-level DP} in FL, which ensures all the training data of a single client are protected. This also fits in the FL setting where each client computes a single model from all its local data.
Assuming a trusted central server, a straightforward idea is to apply DP into the aggregation of model updates for participating clients and hide any client's influence on the model update, at the server. DP-SGD can be adapted to both FedAvg and FedSGD, which forms two DP variants, DP-FedAvg and DP-FedSGD~\cite{mcmahan2018learning}. In a high-level, they work as follows: 1) sampling a group of clients to train local models with total data; 2) clipping the model updates of clients to bound the norm of the total updates; 3) averaging the clipped updates; and 4) adding calibrated Gaussian noise to the average update. 
The privacy amplification via subsampling and moment accountant still apply to compose the privacy loss \cite{geyer2017differentially}. 
However, when providing formal DP guarantee, a particular attention should be paid to a client dropout issue, which may violate the uniform sampling assumption.
Fortunately, recent studies show the possibilities of addressing in theory or bypassing with new framework.
 Despite the existence of noise in both the intermediate model updates and final model, their privacy guarantees are much different as being quantified from different views.

\subsubsection{FL with Local DP}\vspace{3mm}
LDP implemented on local models can defend against untrusted server or other clients. Related studies can be categorized into two lines based on the FL architecture.
 
\textcolor{black}{\textbf{Noise before aggregation}.}
Considering an untrusted central server in practice, LDP can be applied to perturb gradients or model updates for individual client in each iterate. 
A simple approach is to add Gaussian noise to individuals' updates before uploading, which is also known as noising before model aggregation FL~\cite{wei2019performance}. For example, DP-FedSGD or DP-FedAvg can be further adapted into the LDP setting by offloading Gaussian noise addition to the clients' side. Since the summation of multiple Gaussian noises still follows a Gaussian distribution, both the privacy loss at individual clients and the central server can be tracked simultaneously. FL algorithms with LDP, e.g., LDP-FedSGD, face the critical problem of the dimension dependency of communication and privacy. Besides communication overheads, given privacy parameter, the noise needed is substantially proportional to the dimension of model parameter vector. Through selecting a fraction of important dimensions, both noise variance and communication overhead can have a significant reduction~\cite{liu2020fedsel}. Therefore, dimension reduction is commonly used for large models. For instance, updated gradients can be sampled in a subset to reduce communication and truncated in value to compress the noise variance \cite{shokri2015privacy}.

\textcolor{black}{\textbf{Blind flooding with noise}.}
FL can be also implemented in a fully decentralized form without any central entity, thus avoiding a single point failure and improving efficiency for heterogeneous systems.
Its main feature is using peer-to-peer (P2P) communications other than a client-server architecture. A reasonable way to ensure model convergence with full information is to broadcast parameters to close neighbors, which, informally, faces even higher privacy risk than an untrusted server. Moreover, in some opportunistic networks (e.g., mobile crowd sensing or autonomous vehicle networks), the communication topology may be even time-varying and clients may meet unfamiliar neighbors frequently. 
In such a case, LDP is necessary and effective to preserve the privacy of exchanged messages among individual clients. This lead to the problem of decentralized optimization with LDP, which aims to ensure model convergence over a sparse P2P network with noisy local models. However, lacking of a coordinating server, autonomous clients often have to adopt an asynchronous update pattern, which brings new challenges to the decentralized optimization in practice. Nonetheless, it has demonstrated that a differentially private asynchronous decentralized parallel SGD can converge at the same optimal rate as SGD, and have a comparable model utility as the synchronous mode while achieving relatively higher efficiency~\cite{xu2021dp}. 

\subsubsection{FL with Distributed DP}\label{subsubsection: DDP}\vspace{3mm}
As discussed before, DDP can bridge the utility-trust gap between LDP and CDP while eliminating the assumption of trusted server via two cryptographic techniques.
 
\textcolor{black}{\textbf{Privacy amplification by shuffling}}
A line of DDP studies for FL concentrate on the aforementioned secure shuffling technique, which offers amplification of privacy-utility tradeoff via additional anonymization for DP. Before forwarding to the untrusted server, locally perturbed models with minor noise are first permuted randomly to eliminate their client identities by one or more trusted (i.e., secure) shufflers, which can be implemented as a trusted proxy or by delicate cryptographic primitives.
By devising the classic \textit{encoder-shuffler-aggregator} (ESA) framework for adapting FL, LDP-SGD adapted with secure shuffling can achieve both strong iteration-level LDP and good overall CDP for final model, without noticeable accuracy loss~\cite{erlingsson2020encode}. 
For high-dimensional parameters in deep models, shuffling the client identities only may still suffer from linkage attack from side channels. A solution is to split parameter vector and then shuffle the dividends to enhance anonymity~\cite{ijcai2021-217}. 
To further trade off between privacy and utility, subsampling is also an important direction, which should consider the dimension importance~\cite{liu2020flame}. 
Reckoning the benefits of Renyi DP (RDP) and its stronger composition of privacy loss, beyond exploring RDP of subsampled mechanism, a natural extension is to further analyze and exploit RDP and RDP composition in the shuffled model~\cite{10.1145/3460120.3484794}.

\textcolor{black}{\textbf{Secure aggregation of small noises}.} 
Secure aggregation protocols in \cite{bonawitz2017practical} overcomes the practical issue of random client dropouts in cross-device FL, paving the way for FL with DDP via secure aggregation.
However, such protocols often involve modular arithmetic, requiring the quantization of communicating contents (or discrete-valued inputs)for acceptable complexity.
Then, the noise for privacy protection of local models should be also generated in discrete value.
One solution is to generate and add minor discrete noise to the discretized parameters of individual clients before secure aggregation while outputting the aggregate parameters with moderate noise equivalent to the CDP model. 
Binomial or Poisson distribution can approach a similar tradeoff between the utility and privacy of the Gaussian mechanism~\cite{agarwal2018cpsgd}, which however does not achieve RDP or enjoy the state-of-the-art composition and amplification. Simply using discrete Gaussian noise can yield RDP with sharp composition and subsampling-based amplification~\cite{kairouz2021distributed}, but relies on an uncommon sampling mechanism when implementing in software packages. Besides, the summation of discrete Gaussian is not closed and may cause privacy degradation. 
Recently, Skellam mechanism can generate noise distributed according to the differences of two independent Poisson random variables~\cite{agarwal2021skellam}. Skellam noise is closed under summation and can leverage the common Poisson sampling tools to get privacy amplification and sharper RDP bound in theory. 
However, it remains an important problem to develop a practical protocol for production-level FL systems.

\subsubsection{Platforms and Tools for FL with DP}\vspace{3mm}
\textcolor{black}{Towards usable FL with DP, many software frameworks and platforms have been developed to support research-oriented simulations or production-oriented applications.} 
For private deep learning, PySyft\footnote{\small \url{https://github.com/OpenMined/PySyft}} is a Python library that supports FL and DP, and decouples model training from private data. 
\textcolor{black}{Its current version mainly focuses on SMC and HE other than DP implementation.} 
Dedicating to fair evaluation of FL algorithms for the research community, FedML\footnote{\small \url{https://github.com/FedML-AI}} develops an open research library and standardized benchmark with diverse FL paradigms and configurations. The current version only integrates weak DP but provides low-level APIs for security primitives.
Similarly, by providing a high-level interface, PaddleFL\footnote{\small \url{https://github.com/PaddlePaddle/PaddleFL}} supports FL model development with DP and offers a baseline DP-SGD implementation. 
Furthermore, despite the consideration of practical FL settings and recognition of privacy issues, other FL frameworks like FATE\footnote{\small \url{https://github.com/FederatedAI/FATE}} and LEAF\footnote{\small \url{https://github.com/TalwalkarLab/leaf}} still lack deep and flexible supports for DP implementation.
Recently, Sherpa.ai FL developed a unified framework for FL with DP, featuring comprehensive support for DP mechanisms and optimization techniques~\cite{rodriguez2020federated}. Nevertheless, it mainly offers algorithm-level optimization and does not consider practical system implementation.
TensorFlow includes DP and FL implementations in its libraries TensorFlow Privacy and TensorFlow Federated\footnote{\small \url{https://www.tensorflow.org/federated}}, respectively. Both libraries integrate seamlessly with existing TensorFlow models and allow training personalized models with DP. However, its integrated DP mechanisms are relatively fixed in design and do not support customized and flexible optimization. 
Opacus\footnote{\small \url{https://opacus.ai}} is a scalable and efficient library for PyTorch model training with DP. 
It introduces an abstraction of privacy engine that attaches to the standard PyTorch optimizer, which makes DP-SGD implementation much easier without explicitly calling low-level APIs. Beyond ML in PyTorch, it can be easily used in PySyft FL workflows to implement FL with DP.

\subsection{Improving Model Utility for FL with DP}\vspace{3mm}
Existing work underpins the baseline frameworks of FL with DP. 
Aiming at usable FL with DP, 
it is essential to pursue a better tradeoff between model utility and privacy. 
By reviewing common techniques in the fields of DP, ML, and FL, some optimization principles are summarized below.

\subsubsection{Optimization from the perspective of DP}\vspace{3mm}

To seek better tradeoff, there are two directions: reducing unnecessary noise addition, and tracking privacy loss tightly.

\textbf{Clipping bound estimation}: Sensitivity calibration determines the proper noise amplitude by correctly bounding the sensitivity value, is crucial for minimizing the noise variance while guaranteeing certain DP. As mentioned before, a common practice in DP-SGD, thus also in SGD-based FL with DP, is to bound the gradient sensitivity by gradient clipping and then add noise accordingly~\cite{abadi2016deep}. However, an underestimated clipping threshold may cause gradient bias and even model divergence while an overestimated one results in excessive noise addition. Thus, it is important to understand the impact of gradient clipping and dynamically identify the proper clipping bounds during training~\cite{chen2020understanding}. For instance, adaptive gradient clipping via divergence analysis or heuristic estimation, can provably or empirically reduce noise and produces models with higher utility~\cite{pichapati2019adaclip,andrew2019differentially}.

\begin{figure*}
	\centering
	{
		\includegraphics[width=0.7\textwidth, height=175pt]{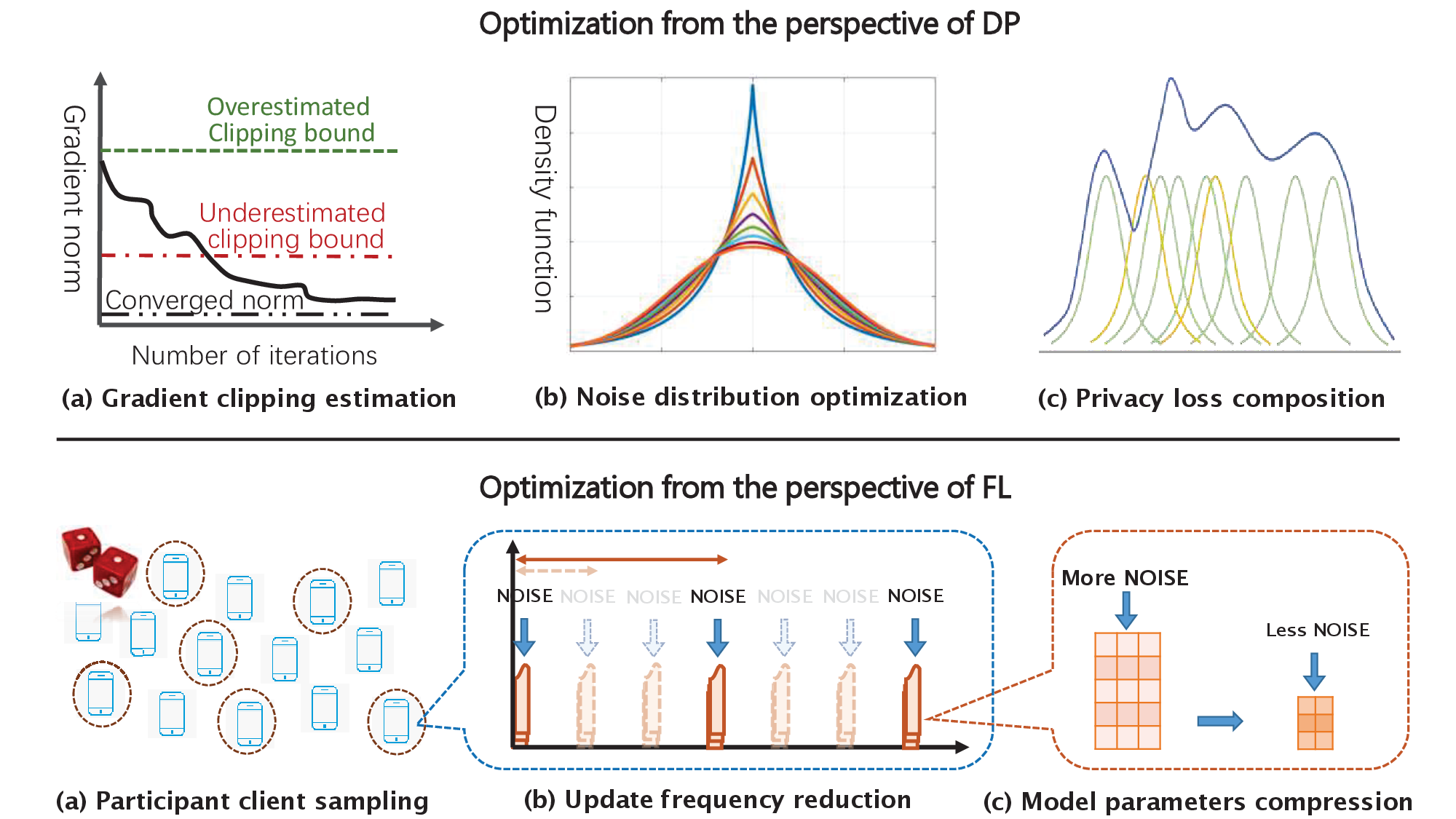}
	}
	\center\caption{Illustration of utility optimization techniques \label{fig: optimization}}
\end{figure*}

\textbf{Noise distribution optimization}: It aims to reduce noise variance by reshaping the noise distribution, thus decreasing unnecessary noise addition in DP.
It has been invested with lots of efforts. For instance, in traditional DP research, some discrete noise distribution and stair-case noise distribution via segmentation techniques have been used in DP algorithms to lessen the necessary noise scale while meeting the DP requirement. In fact, both Lapalce and Gaussian noise for DP are only some instances in a family of the whole distribution space satisfying DP definitions (as shown in Fig.~\ref{fig: optimization}). 
Besides, to incorporating encryption primitives with less overheads, the discretization and quantization of data contents also require the same processing of noise generation for LDP and DDP. 

\textbf{Privacy loss composition}: The composition property of DP allows building complex FL models with DP primitives while composing privacy loss.
Traditionally, both sequential and advanced compositions offer fairly loose bounds.
Moment account analyzes a detailed distribution of the composed privacy loss variable and derives a much tighter bound with higher-order moments. 
It shows acceptable utility with quite small privacy loss for DP-SGD via using amplification techniques~\cite{abadi2016deep}\cite{geyer2017differentially}. 

Privacy loss composition contributes to the optimization of privacy/utility tradeoff by tightly tracking privacy loss for multiple independent noise addition across DP mechanisms~\cite{zhu2019poission}. A relevant but opposite angle is to fix privacy budget and add correlated noises via wise budget division. 
For instance, classic tree aggregation techniques add correlated noises rather than independent ones for repeated computations, which can get high utility while guaranteeing given DP. Inspired by the idea, an amplification-free algorithm adds correlated noise to the accumulation of mini-batch gradients, which achieves a nice tradeoff for DP-SGD without any amplification technique (and no uniform sampling and shuffling requirement)~\cite{kairouz2021practical}.

\textbf{Intrinsic DP computation}:
Many studies have shown that noise-free DP can be achieved by leveraging the inherent randomness of certain models or algorithms for model training, instead of using additional techniques or system components. Being aware of the intrinsic DP level, the designer or developer can save up much budget and add few noise, thus gaining utility without privacy degradation. For instance, by mapping the sampling process to an equivalent exponential mechanism, intrinsic DP in graph models can be effectively measured and leveraged in DP algorithm design. A novel federated model distillation framework can provide provable noise-free DP via random data sampling~\cite{sun2020federated}. It has also been proved that data sketching for communication reduction in FL guarantees DP inherently~\cite{li2019privacy}.
Nonetheless, the intrinsic privacy is not very common and only exists in certain models or algorithms.

\subsubsection{Optimization from the Perspective of FL}\vspace{3mm}
Massive FL clients and the pervasively spatiotemporal sparsity of model parameters offer the chance to extract acceptable utility without significantly harming the privacy.

\textbf{Updating frequency reduction}:
DP enhanced FL suffers from noise accumulation during excessive training epochs. For communication efficiency, too many training epochs also require much network bandwidth.
Therefore, it is highly desirable to reduce the model update frequency. 
Compared with FedSGD, FedAvg allows clients to perform multiple local updates before aggregation, thus reducing global update frequency~\cite{mcmahan2017communication}.
A similar technique has been widely adopted in the DP applications with dynamic datasets or time-series data.
For instance, the data curator publishes perturbed data with DP noise at the timestamps with frequent changes while releasing approximate data without privacy budget consumption at non-changing timestamps.

\textbf{Model parameters compression}:
Like the issue of frequent parameters updating, a long parameter vector heavily consumes the privacy budget (or incurs much noise with the fixed budget) and burdens the limited communication channel. 
To this end, many aforementioned model compression approaches, including parameter filtering, low-rank approximation, random projection, gradient quantization, compressive sensing, etc. have been proposed for deep learning models. \textcolor{black}{For instance, similar studies include sampling and truncating a subset of gradient parameters in FL with CDP~\cite{shokri2015privacy}, selecting top-K dimensions with large contributions in FL with LDP~\cite{liu2020fedsel}, sampling dimensions in FL with DDP~\cite{liu2020flame}.} 
All these methods manage to empirically reduce both the communication bandwidth consumption and noise variance. However, lossy compression techniques, on the one hand, can effectively improve model utility via reducing the DP noise; on the other hand, they may lead to utility loss as some parameter information is eliminated. An immediate question is how to find the optimal compression rate for achieve best utility privacy tradeoff.

\textbf{Participating clients sampling}:
Besides reducing the update frequency and size of parameters, sampling the clients participating in DP-based FL training is also a promising approach to save privacy budget, communication overhead, and energy consumption. The rationale behind this approach comes from the amplification effect of sampling for DP, in which, by randomly sampling the DP protected FL clients in training epochs, much stronger privacy protection can be achieved while
minimizing the average consumption in communication and computation as well as privacy. 
However, in practical cross-device FL, the set of available clients is usually dynamic without prior knowledge of the population. Moreover, as will discuss later, participating clients may drop out randomly. These issues make the assumption of uniform sampling unrealistic and cause severe challenges for gaining privacy-utility tradeoffs \cite{balcer2021connecting}.

\section{Challenges and Discussions}\vspace{3mm}

\textcolor{black}{Despite the great potential and opportunities of DP enhanced FL, there are still challenges in achieving usable FL with DP guarantee in emerging applications.}

\textcolor{black}{
$\bullet$~\textbf{Vertical/Transfer federation}:  
FL can be also categorized according to different data partition strategies.
The above-discussed FL in the generic form, mainly considers the horizontal data partition where each client holds a set of samples with the same feature space. Now, vertical FL where each party holds different features of the same set of samples has gained increasing attention~\cite{fink2021artificial}. However, many existing studies on VFL are based on SMC for protecting confidentiality without considering privacy leakage in the final results. To achieve provable resistance to membership inference or reconstruction attacks, DP must be employed for safeguarding VFL. But it is more challenging than HFL because of two reasons. One is that the VFL algorithm design varies for different tasks and models and often requires case-by-case development. Another is the correlations among distributed attributes are more difficult to identify without spreading individual information to other parties. Besides the vertical federation, there are also scenarios where different parties may hold datasets with non-overlapping features and users. Federated transfer learning (FTL) can eliminate the shifts of feature spaces in this scenario by combining FL and domain adaptation. However, similar to VFL, achieving DP for FTL is still challenging as the gradient of individual instances has to be exchanged between participants. }

\textcolor{black}{
$\bullet$~\textbf{Large language models}: 
With the emergence of large language models (LLMs) like ChatGPT, both FL and DP have begun to demonstrate a promising future in fine-tuning LLMs, while preserving privacy with respect to the private domain data. However, these LLMs often have several billions to hundreds of billions of parameters.
When applying DP and FL to LLMs, there will be multiple challenges concerning the huge number of parameters beyond the extra communication and computation burdens on resource-constrained participants.
Regardless of the DP model, the total amount of privacy noise has to be proportional to the number of parameters for enforcing DP on models, which would lead to huge utility loss.
Besides, the fine-tuning of pre-trained LLMs is also different from conventional model training. 
The theoretical privacy guarantee in ML (e.g., DP-SGD) often assumes models are learned from scratch with many training iterations, instead of a fine-tuning mode with much fewer iterations. 
Therefore, it is necessary to investigate new frameworks for applying both DP and FL and develop new theories for proper privacy guarantees in LLMs.}

\textcolor{black}{$\bullet$~\textbf{FL over streams}:
In many realistic scenarios, training data are continuously generated in the form of streams at distributed clients. In such cases, FL systems have to conduct repetitive analyses on distributed streams. By inheriting online machine learning (OL), online federated learning can be naturally derived to avoid retraining models from scratch each time a new data fragment comes. However, achieving DP for OFL brings multiple challenges. The first is how to define privacy in the OFL setting, as the general DP notion works for static datasets only. Although existing privacy notions for data streams and FL seem to apply here, they still need to be clarified and formulated rigorously in the OFL setting. The second is the efficient algorithm. Taking the event-level LDP (i.e., ensuring $\epsilon$-LDP at each time instance) as an example, frequent uploading of local model updates accumulates huge communication costs and great utility loss, as the noise is proportional to the size of communication data. How to achieve communication and privacy efficiency without degrading overall model performance is thus an important but unsolved research problem.}

\textcolor{black}{Apart from adapting to the new settings, building usable DP-enhanced FL systems still needs to improve its robustness, consider fairness, and allow the data to be forgotten.}

\textcolor{black}{$\bullet$~\textbf{Robustness}.
A robust FL system should be resilient to various failures and attacks caused by misbehaved participants.
Due to limited capabilities (e.g., battery limit), FL clients (e.g., smartphones) may drop out of FL training at any time unexpectedly. 
The random client dropouts bring severe challenges to the practical design of differentially private FL. Except for requiring a more sophisticated design of secure aggregation protocols~\cite{bonawitz2017practical}, some important assumptions may no longer hold for correctly measuring DP in FL. For instance, the DP amplification via shuffling and subsampling both rely on the assumption of clients correctly following the protocol.
Despite recent progress in theory~\cite{bonawitz2017practical,balle2020privacy}, building practical FL systems while addressing the above impacts simultaneously is still challenging. 
Beyond robustness to dropouts of unintended client failure, defending against robustness attacks (e.g., model poisoning for Byzantine and backdoor attacks) mounted by malicious participants is much more challenging~\cite{bagdasaryan2020backdoor}. 
%There are multiple reasons for this.
Specifically, both data heterogeneity and model privacy protection in FL would prevent the server from accurately detecting anomalies and tracking specific participants.}

$\bullet$~\textbf{Fairness}.
Privacy protection is only the first step to encouraging data sharing among a large population. 
Fairness enforcement helps to mitigate the unintended bias on individuals with heterogeneous data. 
However, the dilemma is that DP aims to obscure identifiable attributes while fairness requires the knowledge of individuals' sensitive attribute values to avoid biased results.
The gradient clipping and noise addition in DP can exacerbate the unfairness by decreasing the accuracy of the model over underrepresented classes and subgroups. 
So, the general tension between privacy and fairness calls for ethic-aware FL that respects both issues.
Meanwhile, gradient clipping and noise addition can also enhance the robustness to some extent, as discussed above.
This is also consistent \textcolor{black}{with} the conclusion that there is a tension between fairness and robustness in FL~\cite{li2021ditto}.
The constraints of fairness and robustness compete with each other, as robustness enhancement demands filtering out informative updates with significant model differences. 
Therefore, there is a subtle relationship between privacy, fairness, and robustness in FL. 
While existing studies concentrate on each two of them separately, it would be of significance to unify the interplay of the three simultaneously. 

$\bullet$~%\textbf{Federated unlearning}.
\textbf{Privacy right to be forgotten}. 
The rights of privacy include the ``right to be forgotten'', i.e., users can opt out of private data contribution without leaving any trace. 
As ML models memorize much specific information about training samples, to ensure a specific private sample is totally forgotten, the concept of machine unlearning is proposed to eliminate its influence on trained models. 
However, on the one hand, machine unlearning in the context of FL, i.e., federated unlearning, faces distinct challenges. Specifically, it is much harder to erase the influence of a client's data, as the global model iteratively carries on all participating clients' information~\cite{liu2021federaser}. 
A straightforward idea for resolving the problem is recording historical parameter updates of clients at the server, which may cause significant complexity.
On the other hand, existing machine unlearning has been demonstrated to leak privacy by observing the differences between the original and unlearned models~\cite{chen2021machine}.
DP seems to be one of the promising countermeasures. Therefore, it remains an open question about how to realize efficient and privacy-preserving solutions for federated unlearning.

\section{Conclusion}\vspace{3mm}
\label{sec:conclusion}
\textcolor{black}{With both privacy awareness and regulatory compliance, the meeting of FL and DP, will promote the development of artificial intelligence by unblocking the bottle-necking problem of large-scale ML . 
The article presents a comprehensive overview of the developments, a clear categorization of current advances, and high-level perspectives on the utility optimization principles of FL with DP. This review aims to help the community to better understand the achievements in different ways of combining FL with DP, and the challenges of usable FL with rigorous privacy guarantees. Although FL and DP are increasingly promising in safeguarding private data in the AI era, their combination still faces severe challenges in emerging AI applications. Also, they need further consideration and improvements on other practical issues.}

%\section*{Acknowledgement}
%	This work was supported in part by the National Key Research and Development Program of China under Grant 2020YFA0713900; in part by the National Natural Science Foundation of China under Grants 61772410, 62172329, and U21A6005. 

\bibliographystyle{abbrv}
\bibliography{FLandDP}

\begin{thebibliography}{10}

\bibitem{abadi2016deep}
M.~Abadi, A.~Chu, I.~Goodfellow, H.~B. McMahan, I.~Mironov, K.~Talwar, and
  L.~Zhang.
\newblock Deep learning with differential privacy.
\newblock In {\em Proc. of ACM CCS}, pages 308--318, 2016.

\bibitem{agarwal2021skellam}
N.~Agarwal, P.~Kairouz, and Z.~Liu.
\newblock The skellam mechanism for differentially private federated learning.
\newblock In {\em Proc. of NeurIPS}, 2021.

\bibitem{agarwal2018cpsgd}
N.~Agarwal, A.~T. Suresh, F.~X.~X. Yu, S.~Kumar, and B.~McMahan.
\newblock cpsgd: Communication-efficient and differentially-private distributed
  sgd.
\newblock In {\em Proc. of NeurIPs}, pages 7564--7575, 2018.

\bibitem{andrew2019differentially}
G.~Andrew, O.~Thakkar, H.~B. McMahan, and S.~Ramaswamy.
\newblock Differentially private learning with adaptive clipping.
\newblock {\em arXiv:1905.03871}, 2019.

\bibitem{bagdasaryan2020backdoor}
E.~Bagdasaryan, A.~Veit, Y.~Hua, D.~Estrin, and V.~Shmatikov.
\newblock How to backdoor federated learning.
\newblock In {\em Proc. of AISTATS}, pages 2938--2948, 2020.

\bibitem{balcer2021connecting}
V.~Balcer, A.~Cheu, M.~Joseph, and J.~Mao.
\newblock Connecting robust shuffle privacy and pan-privacy.
\newblock In {\em Proc. of ACM-SIAM SODA}, pages 2384--2403, 2021.

\bibitem{balle2020privacy}
B.~Balle, P.~Kairouz, B.~McMahan, O.~D. Thakkar, and A.~Thakurta.
\newblock Privacy amplification via random check-ins.
\newblock {\em Proc. of NeurIPS}, 33, 2020.

\bibitem{bater2018shrinkwrap}
J.~Bater, X.~He, W.~Ehrich, A.~Machanavajjhala, and J.~Rogers.
\newblock Shrinkwrap: efficient sql query processing in differentially private
  data federations.
\newblock {\em Proc. VLDB Endow.}, 12(3):307--320, 2018.

\bibitem{bonawitz2017practical}
K.~Bonawitz, V.~Ivanov, B.~Kreuter, A.~Marcedone, H.~B. McMahan, S.~Patel,
  D.~Ramage, A.~Segal, and K.~Seth.
\newblock Practical secure aggregation for privacy-preserving machine learning.
\newblock In {\em Proc. of ACM CCS}, pages 1175--1191, 2017.

\bibitem{chaudhuri2009privacy}
K.~Chaudhuri and C.~Monteleoni.
\newblock Privacy-preserving logistic regression.
\newblock In {\em Proc. of NeurIPS}, pages 289--296, 2009.

\bibitem{chen2021machine}
M.~Chen, Z.~Zhang, T.~Wang, M.~Backes, M.~Humbert, and Y.~Zhang.
\newblock When machine unlearning jeopardizes privacy.
\newblock In {\em Proc. of ACM CCS}, pages 896--911, 2021.

\bibitem{chen2020understanding}
X.~Chen, S.~Z. Wu, and M.~Hong.
\newblock Understanding gradient clipping in private sgd: a geometric
  perspective.
\newblock {\em Proc. of NeurIPS}, 33, 2020.

\bibitem{cheng2020federated}
Y.~Cheng, Y.~Liu, T.~Chen, and Q.~Yang.
\newblock Federated learning for privacy-preserving ai.
\newblock {\em Communications of the ACM}, 63(12):33--36, 2020.

\bibitem{cheu2019distributed}
A.~Cheu, A.~Smith, J.~Ullman, D.~Zeber, and M.~Zhilyaev.
\newblock Distributed differential privacy via shuffling.
\newblock In {\em Proc. of Eurocrypt}, pages 375--403, 2019.

\bibitem{duchi2014privacy}
J.~C. Duchi, M.~I. Jordan, and M.~J. Wainwright.
\newblock Privacy aware learning.
\newblock {\em J. ACM}, 61(6):1--57, 2014.

\bibitem{Dwork2011Firm}
C.~Dwork.
\newblock A firm foundation for private data analysis.
\newblock {\em Comm. ACM}, 54(1):86--95, 2011.

\bibitem{dwork2014algorithmic}
C.~Dwork, A.~Roth, et~al.
\newblock The algorithmic foundations of differential privacy.
\newblock {\em Foundations and Trends{\textregistered} in Theoretical Computer
  Science}, 9(3--4):211--407, 2014.

\bibitem{erlingsson2020encode}
{\'U}.~Erlingsson, V.~Feldman, I.~Mironov, A.~Raghunathan, S.~Song, K.~Talwar,
  and A.~Thakurta.
\newblock Encode, shuffle, analyze privacy revisited: Formalizations and
  empirical evaluation.
\newblock {\em arXiv:2001.03618}, 2020.

\bibitem{farokhi2020distributionally}
F.~Farokhi.
\newblock Distributionally-robust machine learning using locally
  differentially-private data.
\newblock {\em arXiv:2006.13488}, 2020.

\bibitem{fink2021artificial}
O.~Fink, T.~Netland, and S.~Feuerriegelc.
\newblock Artificial intelligence across company borders.
\newblock {\em Communications of the ACM}, 65(1):34--36, 2021.

\bibitem{fredrikson2015inversion}
M.~Fredrikson, S.~Jha, and T.~Ristenpart.
\newblock Model inversion attacks that exploit confidence information and basic
  countermeasures.
\newblock In {\em Proc. of ACM CCS}, pages 1322--1333, 2015.

\bibitem{geyer2017differentially}
R.~C. Geyer, T.~Klein, and M.~Nabi.
\newblock Differentially private federated learning: {A} client level
  perspective.
\newblock In {\em Proc. of NeurIPs}, 2017.

\bibitem{10.1145/3460120.3484794}
A.~M. Girgis, D.~Data, S.~Diggavi, A.~T. Suresh, and P.~Kairouz.
\newblock On the r\'{e}nyi differential privacy of the shuffle model.
\newblock In {\em Proc. of ACM CCS}, page 2321–2341, 2021.

\bibitem{hitaj2017deep}
B.~Hitaj, G.~Ateniese, and F.~Perez-Cruz.
\newblock Deep models under the gan: information leakage from collaborative
  deep learning.
\newblock In {\em Proc. of ACM CCS}, pages 603--618, 2017.

\bibitem{hyland2019intrinsic}
S.~L. Hyland and S.~Tople.
\newblock On the intrinsic privacy of stochastic gradient descent.
\newblock {\em arXiv:1912.02919}, 2019.

\bibitem{ji2014differential}
Z.~Ji, Z.~C. Lipton, and C.~Elkan.
\newblock Differential privacy and machine learning: a survey and review.
\newblock {\em arXiv:1412.7584}, 2014.

\bibitem{kairouz2021distributed}
P.~Kairouz, Z.~Liu, and T.~Steinke.
\newblock The distributed discrete gaussian mechanism for federated learning
  with secure aggregation.
\newblock {\em arXiv:2102.06387}, 2021.

\bibitem{kairouz2021practical}
P.~Kairouz, B.~McMahan, S.~Song, O.~Thakkar, A.~Thakurta, and Z.~Xu.
\newblock Practical and private (deep) learning without sampling or shuffling.
\newblock {\em arXiv:2103.00039}, 2021.

\bibitem{kairouz2019advances}
P.~Kairouz, H.~B. McMahan, B.~Avent, A.~Bellet, M.~Bennis, A.~N. Bhagoji,
  K.~Bonawitz, Z.~Charles, G.~Cormode, R.~Cummings, et~al.
\newblock Advances and open problems in federated learning.
\newblock {\em arXiv:1912.04977}, 2019.

\bibitem{konevcny2016federated}
J.~Konecn{\'{y}}, H.~B. McMahan, F.~X. Yu, P.~Richt{\'{a}}rik, A.~T. Suresh,
  and D.~Bacon.
\newblock Federated learning: Strategies for improving communication
  efficiency.
\newblock {\em CoRR}, abs/1610.05492, 2016.

\bibitem{li2021ditto}
T.~Li, S.~Hu, A.~Beirami, and V.~Smith.
\newblock Ditto: Fair and robust federated learning through personalization.
\newblock In {\em Proc. of ICML}, pages 6357--6368, 2021.

\bibitem{li2019privacy}
T.~Li, Z.~Liu, V.~Sekar, and V.~Smith.
\newblock Privacy for free: Communication-efficient learning with differential
  privacy using sketches.
\newblock {\em arXiv:1911.00972}, 2019.

\bibitem{liu2021federaser}
G.~Liu, X.~Ma, Y.~Yang, C.~Wang, and J.~Liu.
\newblock Federaser: Enabling efficient client-level data removal from
  federated learning models.
\newblock In {\em Proc. of IEEE IWQOS}, pages 1--10, 2021.

\bibitem{liu2020flame}
R.~Liu, Y.~Cao, H.~Chen, R.~Guo, and M.~Yoshikawa.
\newblock Flame: Differentially private federated learning in the shuffle
  model.
\newblock In {\em Proc. of AAAI}, number~10, pages 8688--8696, 2021.

\bibitem{liu2020fedsel}
R.~Liu, Y.~Cao, M.~Yoshikawa, and H.~Chen.
\newblock Fedsel: Federated sgd under local differential privacy with top-k
  dimension selection.
\newblock In {\em Proc. of DASFAA}, pages 485--501, 2020.

\bibitem{mcmahan2017communication}
B.~McMahan, E.~Moore, D.~Ramage, S.~Hampson, and B.~A. y~Arcas.
\newblock Communication-efficient learning of deep networks from decentralized
  data.
\newblock In {\em Proc. of AISTAS}, pages 1273--1282, 2017.

\bibitem{mcmahan2018learning}
H.~B. McMahan, D.~Ramage, K.~Talwar, and L.~Zhang.
\newblock Learning differentially private recurrent language models.
\newblock In {\em Proc. of ICLR}, pages 1--10.

\bibitem{melis2019exploiting}
L.~Melis, C.~Song, E.~De~Cristofaro, and V.~Shmatikov.
\newblock Exploiting unintended feature leakage in collaborative learning.
\newblock In {\em Proc. of IEEE S\&P}, pages 691--706, 2019.

\bibitem{nasr2019comprehensive}
M.~Nasr, R.~Shokri, and A.~Houmansadr.
\newblock Comprehensive privacy analysis of deep learning: Passive and active
  white-box inference attacks against centralized and federated learning.
\newblock In {\em Proc. of IEEE S\&P}, pages 739--753. IEEE, 2019.

\bibitem{pichapati2019adaclip}
V.~Pichapati, A.~T. Suresh, F.~X. Yu, S.~J. Reddi, and S.~Kumar.
\newblock Adaclip: Adaptive clipping for private sgd.
\newblock {\em arXiv:1908.07643}, 2019.

\bibitem{RigakiGarcia-45}
M.~Rigaki and S.~Garcia.
\newblock A survey of privacy attacks in machine learning.
\newblock {\em arXiv:2007.07646}, 2020.

\bibitem{rodriguez2020federated}
N.~Rodr{\'\i}guez-Barroso, G.~Stipcich, D.~Jim{\'e}nez-L{\'o}pez, J.~A.
  Ruiz-Mill{\'a}n, E.~Mart{\'\i}nez-C{\'a}mara, G.~Gonz{\'a}lez-Seco, M.~V.
  Luz{\'o}n, M.~A. Veganzones, and F.~Herrera.
\newblock Federated learning and differential privacy: Software tools analysis,
  the sherpa. ai fl framework and methodological guidelines for preserving data
  privacy.
\newblock {\em Inf. Fusion}, 64:270--292, 2020.

\bibitem{roy2020crypt}
A.~Roy~Chowdhury, C.~Wang, X.~He, A.~Machanavajjhala, and S.~Jha.
\newblock Crypt$\epsilon$: Crypto-assisted differential privacy on untrusted
  servers.
\newblock In {\em Proc. of ACM SIGMOD}, pages 603--619, 2020.

\bibitem{shokri2015privacy}
R.~Shokri and V.~Shmatikov.
\newblock Privacy-preserving deep learning.
\newblock In {\em Proc. of ACM CCS}, pages 1310--1321, 2015.

\bibitem{shokri2017mia}
R.~{Shokri}, M.~{Stronati}, C.~{Song}, and V.~{Shmatikov}.
\newblock Membership inference attacks against machine learning models.
\newblock In {\em Proc. of IEEE S\&P}, pages 3--18, 2017.

\bibitem{sun2020federated}
L.~Sun and L.~Lyu.
\newblock Federated model distillation with noise-free differential privacy.
\newblock {\em arXiv:2009.05537}, 2020.

\bibitem{ijcai2021-217}
L.~Sun, J.~Qian, and X.~Chen.
\newblock Ldp-fl: Practical private aggregation in federated learning with
  local differential privacy.
\newblock In {\em Proc. of IJCAI}, 2021.

\bibitem{wagh2021dp}
S.~Wagh, X.~He, A.~Machanavajjhala, and P.~Mittal.
\newblock Dp-cryptography: marrying differential privacy and cryptography in
  emerging applications.
\newblock {\em Comm. ACM}, 64(2):84--93, 2021.

\bibitem{wang2019beyond}
Z.~Wang, M.~Song, Z.~Zhang, Y.~Song, Q.~Wang, and H.~Qi.
\newblock Beyond inferring class representatives: User-level privacy leakage
  from federated learning.
\newblock In {\em Proc. of IEEE INFOCOM}, pages 2512--2520, 2019.

\bibitem{wei2019performance}
K.~Wei, J.~Li, M.~Ding, C.~Ma, H.~H. Yang, F.~Farokhi, S.~Jin, T.~Q. Quek, and
  H.~V. Poor.
\newblock Federated learning with differential privacy: Algorithms and
  performance analysis.
\newblock {\em IEEE Trans. Inf. Forensics Security}, 15:3454--3469.

\bibitem{xu2021dp}
J.~Xu, W.~Zhang, and F.~Wang.
\newblock A (dp)\^{} 2sgd: Asynchronous decentralized parallel stochastic
  gradient descent with differential privacy.
\newblock {\em IEEE Trans. Pattern Anal. Mach. Intell.}, 2021.

\bibitem{zhang2012functional}
J.~Zhang, Z.~Zhang, X.~Xiao, Y.~Yang, and M.~Winslett.
\newblock Functional mechanism: regression analysis under differential privacy.
\newblock {\em Proc. VLDB Endow.}, 5(11):1364--1375, 2012.

\bibitem{zhu2019deep}
L.~Zhu, Z.~Liu, and S.~Han.
\newblock Deep leakage from gradients.
\newblock In {\em Proc. of NeurIPS}, pages 14774--14784, 2019.

\bibitem{zhu2019poission}
Y.~Zhu and Y.-X. Wang.
\newblock Poission subsampled r{\'e}nyi differential privacy.
\newblock In {\em International Conference on Machine Learning}, pages
  7634--7642. PMLR, 2019.

\end{thebibliography}

\end{document}